\documentclass[a4paper,fleqn]{cas-sc}

\usepackage[numbers, sort&compress]{natbib}
\usepackage{placeins}
\usepackage{amsmath,amssymb,amsfonts}
\usepackage{tikz,tabstackengine}
\usetikzlibrary{calc}
\usetikzlibrary{shapes,shapes.geometric,arrows.meta,fit,calc,positioning,automata,patterns, shapes.multipart, arrows.meta}
\usepackage{booktabs}
\usepackage{multirow}
\usepackage{float}
\usepackage{comment}
\usepackage{graphicx}      
\usepackage{subcaption}
\usepackage{overpic}

\definecolor{darkgreen}{RGB}{61,124,68}
\definecolor{plgreen}{RGB}{27,158,119}
\definecolor{HvColor}{rgb}{0.75,0.10,0.10}
\definecolor{VColor}{rgb}{0.10,0.10,0.10}
\definecolor{SColor}{rgb}{0.55,0.55,0.55}
\definecolor{EColor}{rgb}{0.20,0.60,0.20}
\definecolor{IColor}{rgb}{0.60,0.20,0.60}
\definecolor{JColor}{rgb}{0.20,0.50,0.70}
\definecolor{RColor}{rgb}{0.85,0.70,0.15}
\def\tsc#1{\csdef{#1}{\textsc{\lowercase{#1}}\xspace}}
\tsc{TH}
\tsc{AP}

\newtheorem{theorem}{Theorem}

\begin{document}
\let\WriteBookmarks\relax
\def\floatpagepagefraction{1}
\def\textpagefraction{.001}

\shorttitle{Hybrid Optimal Control of Epidemiological Models}    

\shortauthors{T.Halterman and A. Pakniyat}  

\title [mode = title]{Hybrid Optimal Control of Homogeneous
Epidemiological Compartmental Models with Regime Switching}

\author[]{Tyler Halterman}

\credit{Writing - review \& editing, Writing - original draft, Formal analysis, Conceptualization, Software}

\cormark[1]

\ead{twhalterman@crimson.ua.edu}

\affiliation[]{
organization={M$^{3}$AC (Multi-Modal Multi-Agent Control) Lab, University of Alabama},
city={Tuscaloosa},
state={AL},
country={USA}}

\author[]{Ali Pakniyat}

\credit{Writing - review \& editing, Writing - original draft, Formal analysis, Supervision, Conceptualization}

\ead{apakniyat@ua.edu}

\cortext[1]{Corresponding author}

\begin{abstract}
Optimal intervention design is formulated as a hybrid optimal control problem for multi-phase homogeneous epidemiological systems. The system extends a foundational compartmental model through intermediate phases that incorporate work-from-home (WFH) policies and a vaccination protocol, yielding a four-phase hybrid system that captures policy escalation and relaxation. Key characteristics of the resulting hybrid system include (i)~phase-dependent continuous dynamics and running costs that respectively capture distinct disease transmission mechanisms and shifting public health-socioeconomic trade-offs, (ii)~a combination of autonomous and controlled switchings for intervention policies, whose times are co-optimized—whether indirectly via state thresholds or directly as decision variables—alongside continuous inputs to minimize the overall cost, and (iii)~nontrivial state jump maps that govern transitions between phases with differing state and control space dimensions. The Hybrid Minimum Principle (HMP) is invoked to obtain the optimal solutions. Numerical results demonstrate that coordinating WFH policies with vaccination efforts provides improved mitigation of disease spread compared to single-phase policy interventions.
\end{abstract}


\let\printorcid\relax
\maketitle


\section{Introduction}\vspace{6pt}
Mathematical modeling and control of infectious diseases play a central role in public health planning, particularly in the design of intervention strategies such as vaccination, quarantine, and isolation. Compartmental epidemiological models, originating from the seminal work of \citep{Kermack}, provide a systematic framework for describing disease transmission through structured population groups as, e.g., presented in  
\citep{Hethcote}. By partitioning populations into epidemiological states and modeling transitions between them, these models offer both analytical tractability and practical insight into disease dynamics.

During the past century, the classical susceptible–infected–recovered (SIR) formulation has been extensively generalized to incorporate additional compartments, including exposed, vaccinated, quarantined, and asymptomatic populations \citep{Poonia, Safarishahrbijari,Zhao}. These extensions enable more realistic representations of disease progression and intervention strategies. In recent years, adaptations to this fixed model structure have been considered to incorporate stochasticity and heterogeneity \citep{pang2025stochastic}, multiplex networks \citep{multi-plex_networks,rozan2025multiplex}, and finite state graphons for mean-field games \citep{aurell2022finite}. In parallel, optimal control theory has been widely applied to determine time-dependent policies that mitigate epidemic burden while considering resource constraints \citep{Lenhart, Behncke}. Despite these advances, most optimal control formulations assume a single fixed model structure over the entire planning time horizon \citep{ramezanzadeh2026optimal,dellarossa2024optimality}.

However, in practice, epidemic response strategies evolve in stages, transitioning between baseline transmission conditions and more aggressive intervention regimes such as lockdowns, vaccination campaigns, or targeted isolation. These shifts naturally give rise to hybrid dynamical systems in which the governing dynamics, control inputs, and even state dimensions may change over~time.

Hybrid optimal control theory (see, e.g., \cite{BensoussanMenaldiStochastic, BranickyBorkerMitter, DharmattiRamaswamy, Sussmann, GaravelloPiccoli, shaikh2007hybrid, FarzinPECSIAM, PakniyatCaines2014, PakniyatCaines2017b, SanfeliceAltin2026TAC}) provides a rigorous framework for addressing such problems. In particular, the Hybrid Minimum Principle (HMP) presented, e.g., in \citep{Sussmann, GaravelloPiccoli, shaikh2007hybrid, FarzinPECSIAM, PakniyatCaines2014, PakniyatCaines2017b} enables the characterization of optimal inputs and trajectories in systems with switching dynamics, including the determination of optimal switching times and the treatment of autonomous and controlled transitions. The HMP has proven highly effective for solving multi-modal problems in domains ranging from automotive systems \citep{APPEC2017NAHS} and AeroMarine multi-modal drones \citep{YasiniASME2023, YasiniADHS2024} to mathematical finance \citep{DFAPPEC2017CDC}. 

Despite many public health interventions being intrinsically hybrid—driven by abrupt policy enactments and distinct regime dynamics—the rigorous application of hybrid optimal control to epidemiology remains largely unexplored. Existing approaches in epidemiology often focus on modifying control inputs within a single model structure or introducing switching logic without fully exploiting multi-phase system dynamics \citep{Bolzoni,Kantner}. Some applications have resulted in epidemiological formulations that consider a hybrid structure \citep{tran2021optimal, dharmatti2023hybrid}; however, the state and control spaces often remain unchanged between phase transitions.

To overcome these limitations and to explicitly capture the hybrid nature of public health interventions, we formulate a multi-phase epidemiological optimal control problem inspired by workplace responses to infectious disease outbreaks. Specifically, we model an office environment that transitions between operational regimes as infection levels evolve. The system begins with baseline transmission dynamics and undergoes an autonomous transition to a work-from-home (WFH) regime once infections exceed a prescribed threshold. A vaccination protocol phase is subsequently introduced through a controlled switching mechanism, and, as conditions improve, the system autonomously transitions back to baseline dynamics, representing a return-to-office (RTO) phase. Within this setting, we leverage the HMP version in \citep{PakniyatCaines2021, PakniyatCaines2023} to develop a framework that captures these regime transitions through phase-dependent dynamics, control inputs, and cost structures, effectively balancing disease mitigation, intervention costs, and operational objectives in a dynamically evolving environment.

The rest of the paper is organized as follows: Section \ref{sec:hybrid_systems_structure} presents the problem formulation, introduces the compartmental models, and the structure of the hybrid system. Section \ref{sec:min_control_effort} presents and applies the HMP to find the optimal control inputs and switching times. Section \ref{sec:num_sim} provides numerical simulations that demonstrate the effectiveness of the proposed framework. Section \ref{sec:conclusion} concludes the paper and discusses future research directions.

\section{Hybrid Systems Modeling of the Epidemiological System}
\label{sec:hybrid_systems_structure}\vspace{6pt}
\subsection{Hybrid Systems Structure}\vspace{6pt}
Following \citep{PakniyatCaines2017b, PakniyatCaines2021, PakniyatCaines2023}, a \textit{hybrid system (structure)} $\mathbb{H}$ is considered as a septuple
\begin{equation}
    \mathbb{H} = \{H, I, \Gamma, A, F, \Xi, \mathcal{M} \},
\end{equation}
where $H := \bigsqcup_{q \in Q} \mathbb{R}^{n_q}$ is the hybrid state space with $|Q| < \infty$; in this paper, $Q = \{q_1, q_2, q_3\} \equiv \{\text{rto}, \text{wfh}, \text{protocol}\}$, with $n_{q_1}=6$, $n_{q_2}=8$, and $n_{q_3}=7$.

$I := \Sigma \times \bigsqcup_{q \in Q} U_{q}$ is the input value set with $|\Sigma| < \infty$, and $U_{q} \subset \mathbb{R}^{m_{q}}$
; where, for the epidemiological system, $\Sigma = \{\sigma_{\text{wfh}},\sigma_{\text{protocol}},\,\sigma_{\text{rto}}\}$, $U_{q_{1}} 
\hspace{-2pt} \overset{\text{\scalebox{.6}{compact}}}{\subseteq} \! \mathbb{R}$, and $U_{q_{2}}, U_{q_{3}} \! \overset{\text{\scalebox{.6}{compact}}}{\subseteq} \! \mathbb{R}^3$ with $\sigma_{\text{wfh}}$ and $\sigma_{\text{rto}}$ corresponding to autonomous switchings and $\sigma_{\text{protocol}}$ corresponding to a controlled switching.

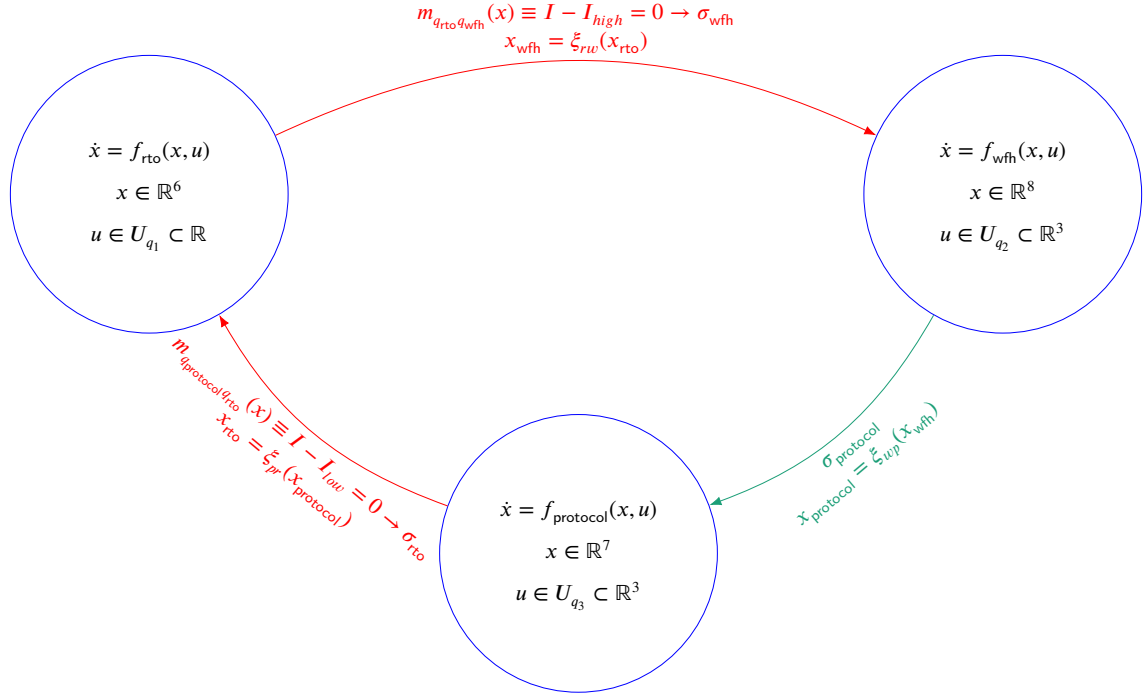
\begin{figure}[]
\centering
\scalebox{0.95}{
\begin{tikzpicture}[node distance=50mm, auto]

\node[state,minimum size=110pt,draw=blue] (q1) {$\begin{array}{c}
\dot{x} = f_{\text{rto}}(x,u)\\[2mm]
x \in \mathbb{R}^{6}\\[2mm]
u \in U_{q_1} \subset \mathbb{R}
\end{array}$};

\node[state,minimum size=110pt,draw=blue] (q2) [right= 80mm of q1] {$\begin{array}{c}
\dot{x} = f_{\text{wfh}}(x,u)\\[2mm]
x \in \mathbb{R}^{8}\\[2mm]
u \in U_{q_2} \subset \mathbb{R}^{3}
\end{array}$};

\node[state,minimum size=110pt,draw=blue] (q3) [below of=q1, xshift=170pt] {$\begin{array}{c}
\dot{x} = f_{\text{protocol}}(x,u)\\[2mm]
x \in \mathbb{R}^{7}\\[2mm]
u \in U_{q_3} \subset \mathbb{R}^{3}
\end{array}$};

\path[-{Latex[scale=1.0]}]

(q1) edge[bend left=25, red]
node[pos=0.50, above, align=center] {
$m_{q_{\text{rto}}q_{\text{wfh}}}(x) \equiv I -  I_{high} = 0 \rightarrow \sigma_{\text{wfh}}$\\
$x_{\text{wfh}}=\xi_{rw}(x_{\text{rto}})$
}
(q2)

(q2) edge[bend left=20, plgreen]
node[pos=0.5,below, sloped, align=center] {
$\sigma_{\text{protocol}}$\\
$x_{\text{protocol}}=\xi_{wp}(x_{\text{wfh}})$
}
(q3)

(q3) edge[bend left=20, red]
node[pos=0.5, below, sloped, align=center] {
$m_{q_{\text{protocol}}q_{\text{rto}}}(x) \equiv I - I_{low} = 0 \rightarrow \sigma_{\text{rto}}$\\
$x_{\text{rto}}=\xi_{pr}(x_{\text{protocol}})$
}
(q1);

\end{tikzpicture}
}
\caption{Hybrid automaton for the multi-phase epidemiological system with {\color{plgreen}controlled switching in green} and {\color{red} autonomous switchings in red}.}
\label{fig:Hybrid_Automaton}
\end{figure}

$\Gamma : H \times \Sigma \to H$ is a \textit{discrete state transition map} which, for the epidemiological system, is displayed as part of Fig.~\ref{fig:Hybrid_Automaton}.

$A : Q \times \Sigma \to Q$ denotes both a deterministic finite automaton and the automaton’s associated transition function on the discrete state space $Q$ and the event set $\Sigma$, as displayed in Fig. \ref{fig:Hybrid_Automaton}.

$F := \{f_q\}_{q \in Q}$ is an indexed collection of \textit{vector fields} whose elements are continuously differentiable functions presented in Section \ref{sec:hybrid_formulation}.

$\Xi : H \times \Sigma \to H$ is a family of continuous state jump transition maps which is presented in Section \ref{sec:hybrid_formulation}.

$\mathcal{M}$ is a collection of switching manifolds whose elements are presented in Section \ref{sec:hybrid_formulation}.

\subsection{Hybrid Dynamics Modeling}
\begin{figure}[]
\centering
\begin{tikzpicture}[
    scale=0.75,             
    transform shape,        
    >=latex,
    node distance=1.8cm,    
    compartment/.style={circle, draw, minimum size=0.75cm, font=\bfseries},
    every path/.style={line width=0.6pt, ->, -{Latex}}
]

\node[compartment, fill=blue!30] (V) {$V$};
\node[compartment, fill=blue!15, below=of V] (S) {$S$};
\node[compartment, fill=green!20, right=of V] (E) {$E$};
\node[compartment, fill=yellow!20, right=of E] (I) {$I$};
\node[compartment, fill=gray!30, above=of I] (J) {$J$};
\node[compartment, fill=orange!30, right=of I] (R) {$R$};

\draw (V) -- node[xshift = -1,above] {$\beta_v V I$} (E);
\draw (S) -- node[left] {$\beta_s S I$} (E);

\draw (E) -- node[above] {$\kappa E$} (I);

\draw[magenta] (I) -- node[right] {$u_jI$} (J);

\draw (I) -- node[above] {$\gamma I$} (R);
\draw (J) -- node[right] {$\delta J$} (R);

\draw (R) to[bend left=30] node[below] {$\omega R$} (S);

\draw[dashed, ->, black] (I) to[bend right=70] ($(V)!0.25!(E)$);
\draw[dashed, ->, black] (I) to[bend left=40] ($(S)!0.2!(E)$);

\end{tikzpicture}

\caption{Transition map for normal / return to office~(RTO), labeled by the discrete state $q_1$, with {\color{magenta} control pathways in magenta}, uncontrolled pathways in black, and dashed arrows indicating state-dependent infection interactions.}
\label{fig:RTO_tikz}
\end{figure}
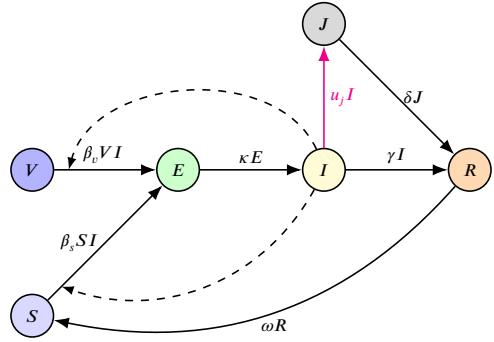

\vspace{6pt}
\label{sec:hybrid_formulation}
The epidemiological dynamics during the normal working and return-to-office (RTO) conditions of the first and fourth phases, as visualized by the compartmental transitions in Fig.~\ref{fig:RTO_tikz}, are governed by
\begin{equation} \label{eq:baseline_statevector_new}
    \dot{x}_{q_1} \equiv \left[\begin{array}{c}
\dot V\\
\dot S\\
\dot E\\
\dot I\\
\dot J\\
\dot R
\end{array}\right]= f_{q_1}(x_{q_1},u) \equiv \left[\begin{array}{c}
-\beta_{v}VI\\
-\beta_{s}SI+\omega R\\
\beta_{v}VI+\beta_{s}SI-\kappa E\\
\kappa E-\gamma I-u_jI\\
u_jI-\delta J\\
\gamma I+\delta J-\omega R
\end{array}\right]
\end{equation}
The initial condition is given by $x_{q_1}(t_0)=x_0$. In this model, the state components $V$ and $S$ denote vaccinated and unvaccinated susceptible populations while $E$, $I$, $J$, and $R$ denote, respectively, the exposed, infectious, quarantined, and recovered populations. The infection process is governed by the transmission coefficients $\beta_v$ and $\beta_s$, reflecting different exposure risks for vaccinated and unvaccinated individuals. The parameter $\kappa$ governs progression from the exposed to the infectious compartment, $\gamma$ denotes recovery from infection, $\delta$ represents removal from quarantine, and $\omega$ captures waning immunity through the transition from the recovered class back to the unvaccinated susceptible class. The control input $u_j$ represents quarantine effort through contact testing and takes values from the set $[0,u_j^{\max}]$, where we consider $u_j^{\max} = 0.04$ for the maximum capacity for intervention intensity.

Once the infected population $I$ has increased to a predefined threshold $I_{high}$, the first RTO phase must end and the WFH phase is enforced. This is represented by~the switching manifold condition $m_{q_1q_2}(x_{q_1}) = x_{q_1}^{(4)} - I_{high} \equiv I - I_{high} =0$. In other words, $t_{s_1}$ marks the instant at which $m_{q_1q_2}\big(x_{q_1} (t_{s_1}-)\big) \equiv \lim_{t\rightarrow t_{s_1}}m_{q_1q_2}\big(x_{q_1} (t)\big) = 0$.

Consequently, the dynamics of the epidemiological system in the second phase, visualized in Fig.~\ref{fig:WFH_tikz}, will change to
\begin{equation} \label{eq:fq2_new}
    \dot{x}_{q_2} \equiv \left[\begin{array}{c}
\dot H_v\\
\dot H_s\\
\dot V\\
\dot S\\
\dot E\\
\dot I\\
\dot J\\
\dot R
\end{array}\right]
=
f_{q_2}(x_{q_2},u)
=
\begin{bmatrix}
u_{\sigma_v}V\\
u_{\sigma_s}S\\
-\beta_v V I - u_{\sigma_v}V\\
-\beta_s S I - u_{\sigma_s}S + \omega R\\
\beta_v V I + \beta_s S I - \kappa E\\
\kappa E - \gamma I - u_jI\\
u_jI - \delta J\\
\gamma I + \delta J - \omega R
\end{bmatrix}
\end{equation}
where $H_v$ and $H_s$ denote vaccinated and unvaccinated individuals assigned to WFH status. This phase extends $q_{1}$ through the introduction of two additional compartments that temporarily remove individuals from in-person exposure pathways. The control inputs $u_{\sigma_v}$ and $u_{\sigma_s}$ govern the rate at which vaccinated and unvaccinated susceptible individuals transition into the WFH classes, respectively. In this phase, the quarantine control $u_j$ takes values from the set $[0,u_j^{max}]$, while the WFH controls, $u_{\sigma_v}$ and $u_{\sigma_s}$, take values from the sets $[0,u_{\sigma_v}^{max}]$ and $[0,u_{\sigma_s}^{max}]$ respectively. We consider $u_j^{\max} = 0.04$, $u_{\sigma_v}^{max} = 0.25$, and $u_{\sigma_s}^{max} = 0.25$  for the maximum capacity for intervention intensity.

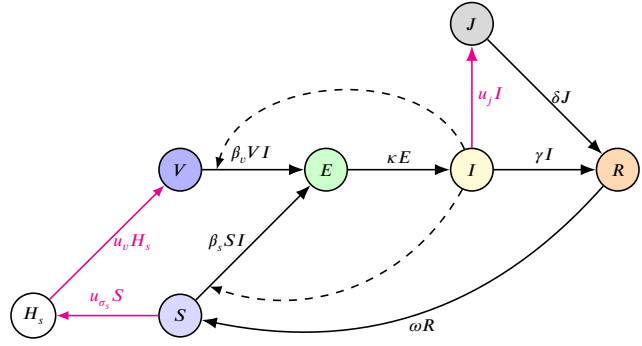
\begin{figure}[]
\centering
\begin{tikzpicture}[
    scale=0.75,             
    transform shape,        
    >=latex,
    node distance=1.8cm,    
    compartment/.style={circle, draw, minimum size=0.75cm, font=\bfseries},
    every path/.style={line width=0.6pt, ->, -{Latex}}
]
\node[compartment, fill=blue!30] (V) {$V$};
\node[compartment, fill=blue!15, below=of V] (S) {$S$};
\node[compartment, fill=green!20, right=of V] (E) {$E$};
\node[compartment, fill=yellow!20, right=of E] (I) {$I$};
\node[compartment, fill=gray!30, above=of I] (J) {$J$};
\node[compartment, fill=orange!30, right=of I] (R) {$R$};
\node[compartment, fill=white!30, left=of S] (Hs) {$H_s$};

\draw (V) -- node[xshift=-1,above] {$\beta_v V I$} (E);
\draw (S) -- node[left] {$\beta_s S I$} (E);

\draw (E) -- node[above] {$\kappa E$} (I);

\draw[magenta] (I) -- node[right] {$u_jI$} (J);

\draw (I) -- node[above] {$\gamma I$} (R);
\draw (J) -- node[right] {$\delta J$} (R);

\draw (R) to[bend left=30] node[below] {$\omega R$} (S);

\draw (S)[->, magenta] -- node[above] {$u_{\sigma_s}S$} (Hs);
\draw (Hs)[->, magenta] -- node[right] {$u_v H_s$} (V);

\draw[dashed, ->, black] (I) to[bend right=70] ($(V)!0.25!(E)$);
\draw[dashed, ->, black] (I) to[bend left=40] ($(S)!0.2!(E)$);

\end{tikzpicture}
\caption{Transition map for vaccination protocol, labeled by the discrete state $q_3$, with {\color{magenta} control pathways in magenta}, uncontrolled pathways in black, and dashed arrows indicating state-dependent infection interactions.}
\label{fig:Protocol_tikz}
\end{figure}

The initial condition of mode $q_2$ is determined from the pre-switching state value in mode $q_1$ according to the jump map
\begin{equation}
x_{q_2}(t_{s_1})
\equiv 
\begin{bmatrix}
H_v(t_{s_1})\\
H_s(t_{s_1})\\
V(t_{s_1})\\
S(t_{s_1})\\
E(t_{s_1})\\
I(t_{s_1})\\
J(t_{s_1})\\
R(t_{s_1})
\end{bmatrix}
=
\xi_{q_1q_2}\left(x_{q_1}(t_{s_1}-)\right) = 
\begin{bmatrix}
0\\
0\\
V(t_{s_1}-)\\
S(t_{s_1}-)\\
E(t_{s_1}-)\\
I(t_{s_1}-)\\
J(t_{s_1}-)\\
R(t_{s_1}-)
\end{bmatrix}
\end{equation}
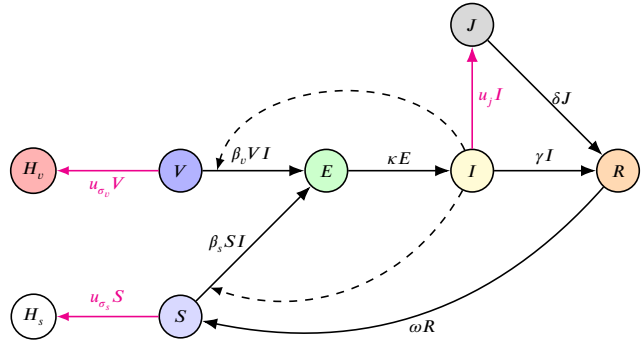
\begin{figure}[]
\centering
\begin{tikzpicture}[
    scale=0.75,             
    transform shape,        
    >=latex,
    node distance=1.8cm,    
    compartment/.style={circle, draw, minimum size=0.75cm, font=\bfseries},
    every path/.style={line width=0.6pt, ->, -{Latex}}
]

\node[compartment, fill=blue!30] (V) {$V$};
\node[compartment, fill=blue!15, below=of V] (S) {$S$};
\node[compartment, fill=green!20, right=of V] (E) {$E$};
\node[compartment, fill=yellow!20, right=of E] (I) {$I$};
\node[compartment, fill=gray!30, above=of I] (J) {$J$};
\node[compartment, fill=orange!30, right=of I] (R) {$R$};
\node[compartment, fill=white!30, left=of S] (Hs) {$H_{s}$};
\node[compartment, fill=red!30, left=of V] (Hv) {$H_v$};

\draw (V) -- node[xshift = -1,above] {$\beta_v V I$} (E);
\draw (S) -- node[left] {$\beta_s S I$} (E);

\draw (E) -- node[above] {$\kappa E$} (I);

\draw[magenta] (I) -- node[right] {$u_jI$} (J);

\draw (I) -- node[above] {$\gamma I$} (R);
\draw (J) -- node[right] {$\delta J$} (R);

\draw (R) to[bend left=30] node[below] {$\omega R$} (S);

\draw (S)[->, magenta] -- node[above] {$u_{\sigma_s}S$} (Hs);
\draw (V)[->, magenta] -- node[below] {$u_{\sigma_v}V$} (Hv);

\draw[dashed, ->, black] (I) to[bend right=70] ($(V)!0.25!(E)$);
\draw[dashed, ->, black] (I) to[bend left=40] ($(S)!0.2!(E)$);

\end{tikzpicture}

\caption{Transition map for work from home (WFH), labeled by the discrete state $q_2$, with {\color{magenta} control pathways in magenta}, uncontrolled pathways in black, and dashed arrows indicating state-dependent infection interactions.}
\label{fig:WFH_tikz}
\end{figure}
The vaccination protocol phase can be initiated at an arbitrary time $t_{s_2}$ during the WFH phase. This marks a controlled switching to the dynamics of the epidemiological system in the third phase, visualized in Fig.~\ref{fig:Protocol_tikz}, expressed as
\begin{equation} \label{eq:fq3_new}
\dot{x}_{q_3} \equiv \left[\begin{array}{c}
\dot H_s\\
\dot V\\
\dot S\\
\dot E\\
\dot I\\
\dot J\\
\dot R
\end{array}\right]
=
f_{q_3}(x_{q_3},u)
=
\begin{bmatrix}
u_{\sigma_s}S - u_vH_s\\
-\beta_v V I + u_vH_s\\
-\beta_s S I - u_{\sigma_s}S + \omega R\\
\beta_v V I + \beta_s S I - \kappa E\\
\kappa E - \gamma I - u_jI\\
u_jI - \delta J\\
\gamma I + \delta J - \omega R
\end{bmatrix}
\end{equation}
In this phase, the vaccinated WFH compartment is removed and the remaining unvaccinated WFH population $H_s$ is retained. The vaccination protocol control $u_v$ transfers individuals from $H_s$ into the vaccinated susceptible class $V$, while the WFH control $u_{\sigma_s}$ continues to move unvaccinated susceptible individuals into $H_s$. During this phase, $u_j$ takes values from the set $[0, u_j^{max}]$, $u_v$ takes values from the set $[0,u_v^{max}]$, and $u_{\sigma_s}$ takes values from the set $[0, u_{\sigma_s}^{max}]$, where $u_j^{max}$ = 0.04, $u_v^{max}$ = 0.05, and $u_{\sigma_s}^{max}$ = 0.1 for the maximum capacity for intervention intensity.

The corresponding jump map for this phase change is described by:
\begin{equation}
x_{q_3}(t_{s_2})
\equiv 
\begin{bmatrix}
H_s(t_{s_2})\\
V(t_{s_2})\\
S(t_{s_2})\\
E(t_{s_2})\\
I(t_{s_2})\\
J(t_{s_2})\\
R(t_{s_2})
\end{bmatrix}
=
\xi_{q_2 q_3}\left(x_{q_2}(t_{s_2}-)\right) =
\begin{bmatrix}
H_s(t_{s_2}-)\\
V(t_{s_2}-) + H_v(t_{s_2}-)\\
S(t_{s_2}-)\\
E(t_{s_2}-)\\
I(t_{s_2}-)\\
J(t_{s_2}-)\\
R(t_{s_2}-)
\end{bmatrix}
\end{equation}

Once the infected population $I$ has reduced to a predefined threshold $I_{low}$, the vaccination protocol phase ends and the final RTO phase begins. This is represented by the switching manifold condition $m_{q_3q_1}(x_{q_3}) = x_{q_3}^{(5)} - I_{low} \equiv I - I_{low} = 0$. In other words, $t_{s_3}$ marks the instant at which $m_{q_3q_1}\big(x_{q_3} (t_{s_3}-)\big) \equiv \lim_{t\rightarrow t_{s_3}}m_{q_3q_1}\big(x_{q_3} (t)\big) = 0$.

This will change the dynamics to \eqref{eq:baseline_statevector_new} subject to the jump map
\begin{equation}
x_{q_1}(t_{s_3})
\equiv 
\begin{bmatrix}
V(t_{s_3})\\
S(t_{s_3})\\
E(t_{s_3})\\
I(t_{s_3})\\
J(t_{s_3})\\
R(t_{s_3})
\end{bmatrix}
=
\xi_{q_3 q_1}\left(x_{q_3}(t_{s_3}-)\right)
=
\begin{bmatrix}
V(t_{s_3}-)\\
S(t_{s_3}-) + H_s(t_{s_3}-)\\
E(t_{s_3}-)\\
I(t_{s_3}-)\\
J(t_{s_3}-)\\
R(t_{s_3}-)
\end{bmatrix}
\end{equation}

\subsection{Associated Hybrid Optimal Control Problem}\vspace{6pt}
The objective of the associated hybrid optimal control problem (HOCP) is to minimize a hybrid cost functional consisting of running costs and a terminal cost:
\begin{equation}
\label{eq:J_new}
\begin{aligned}
J
&=
\int_{0}^{t_{s_1}} \ell_{q_1}^{1}(x_{q_1}(s),u(s))\,ds + \int_{t_{s_{1}}}^{t_{s_2}} \ell_{q_2}(x_{q_2}(s),u(s))\,ds +\int_{t_{s_{2}}}^{t_{s_3}} \ell_{q_3}(x_{q_3}(s),u(s))\,ds \\
&\quad + \int_{t_{s_{3}}}^{t_{f}} \ell_{q_1}^{4}(x_{q_1}(s),u(s))\,ds + \Phi(x(t_f)),
\end{aligned}
\end{equation}
where $\ell_{q_1}^{1}$ and $\ell_{q_1}^{4}$ denote the RTO costs in the first and fourth phases, while $\ell_{q_2}$ and $\ell_{q_3}$ correspond to the WFH and vaccination protocol phases, respectively, and are defined as
\begin{equation}
\begin{aligned}
\ell_{q_1}^{1} &= a_{I_1} I(s) + a_{J_1} J(s) + b_{j_1} u_j(s)^2 \\
\ell_{q_2} &= a_{I_2} I(s) + a_{J_2} J(s) + a_{H_{v_2}} H_v(s) + a_{H_{s_2}} H_s(s) + c_2 + b_{j_2} u_j(s)^2 + b_{\sigma_{v_2}} u_{\sigma_v}(s)^2 + b_{\sigma_{s_2}} u_{\sigma_s}(s)^2 \\
\ell_{q_3} &= a_{I_3} I(s) + a_{J_3} J(s) + a_{H_{s_3}} H_s(s) + c_3 + b_{j_3} u_j(s)^2 + b_{v_3} u_v(s)^2 + b_{\sigma_{s_3}} u_{\sigma_s}(s)^2 \\
\ell_{q_1}^{4} &= a_{E_4} E(s) + a_{I_4} I(s) + a_{J_4} J(s) + b_{j_4} u_j(s)^2.
\end{aligned}
\end{equation}

The running costs $\ell_{q_i}$ are defined as phase-dependent weighted linear combinations of epidemiological states with quadratic penalties on control inputs. This linear–quadratic structure is consistent with standard epidemiological optimal control formulations that balance disease burden and intervention effort \citep{neilan2010optimal}. In particular, all phases penalize the infectious and quarantined populations ($I$, $J$) and include a quadratic cost on the quarantine control $u_j$. 
Additional terms are introduced in specific phases to reflect intervention mechanisms. In the WFH phase, $\ell_{q_2}$ includes linear penalties on the work-from-home populations ($H_v$, $H_s$) and quadratic costs on the controls $u_{\sigma_v}$ and $u_{\sigma_s}$. In the vaccination protocol phase, $\ell_{q_3}$ penalizes $H_s$ and includes quadratic costs on $u_v$ and $u_{\sigma_s}$. The final RTO cost $\ell_{q_1}^{4}$ additionally penalizes the exposed population $E$, promoting suppression of latent infections as normal operations resume. Constant terms in $\ell_{q_2}$ and $\ell_{q_3}$ capture phase-specific baseline operational costs.

 A terminal cost $\Phi(x(t_f))$ penalizes the residual disease burden at the end of the horizon, encouraging convergence to low-prevalence states \citep{liu2023seir_terminal}, defined as
\begin{equation}
\Phi\big(x(t_f)\big) = k_E E(t_f) + k_I I(t_f) + k_J J(t_f).
\end{equation}

\section{Determination of Optimal Control Inputs}\vspace{6pt}
\label{sec:min_control_effort}

In order to determine the optimal inputs, consisting of the optimal continuous inputs $u^*{(\cdot)}$ and the optimal switching times $t^*_{s_i}$, we invoke the Hybrid Minimum Principle (HMP) presented below.

\begin{theorem}{\citep[Theorem 4.1]{PakniyatCaines2021}}\label{thm:HMP}
Define the family of system Hamiltonians as 
\begin{equation}
    H_q(x_q, \lambda_q, u_q, t) = \ell_q(x_q, u_q, t) + \lambda_q^\top f_q(x_q, u_q, t) 
\end{equation}
Then, for an optimal input $u^*{(\cdot)}$ and along the corresponding optimal trajectory $x^*{(\cdot)}$, there exist adjoint processes $\lambda^*{(\cdot)}$  such~that
\begin{equation}
\label{eq.H}
H_{q}\left(x_q^*(t), \lambda_q^*(t), u_q^*(t), t\right) \leq H_{q}\left(x_q^*(t), \lambda_q^*(t), u_q, t\right), \hfill \forall u_q \in U_q, \hspace{10pt}  \forall t \in [t_0,t_f],
\end{equation}
where $x_q^*$ and $\lambda_q^*$ satisfy Hamiltonian canonical equations  
\begin{align}
\label{eq.x*}
    \dot{x}_q^*(t) &= \frac{\partial H_q}{\partial \lambda_q} \big(x_q^*(t), \lambda_q^*(t), u_q^*(t), t\big) \equiv f\big(x_q^*(t),u_q^*(t),t\big)
\\
    \dot{\lambda}_q^*(t) &= - \frac{\partial H_q}{\partial x_q} \big(x_q^*(t), \lambda_q^*(t), u_q^*(t), t\big) \equiv -  \frac{\partial \ell_q}{\partial x}\big(x_q^*(t),u_q^*(t),t\big) - \frac{\partial f_q}{\partial x}\big(x_q^*(t),u_q^*(t),t\big)^\top \lambda_q^*(t)
    \label{eq.la*}
\end{align}
subjected to
\begin{align}
    x^*_{q_1}(t_0) &= x_0  
\\
    x^*_{q_{i+1}}(t_{s_i}) &= \xi_{q_{i} q_{i+1}}\left(x^*_{q_i}(t_{s_i}-)\right)  
\\
\label{eq.b.A}
     \lambda^*_{q_i} (t_{s_i}) &=    \left.\nabla\xi_{q_{i} q_{i+1}}\right|_{x_{q_i}(t_{s_i}-)}^\top \lambda^*_{q_{i+1}} (t_{s_i}\!+) + p_i \left.\nabla m_{q_{i} q_{i+1}}\right|_{x_{q_i}(t_{s_i}-)} 
\\
    \lambda^{*}_{q_1}(t_f) &= \nabla \Phi(x(t_f))  
\end{align}
where $p_i=0$ at the time of a controlled switching and $p_i \in \mathbb{R}$ at the time of an autonomous switching subject to the additional switching manifold condition $m_{q_{i} q_{i+1}}\big(x_{q_i}(t_{s_i}-)\big) = 0$.

Moreover, at the optimal switching times $t_{s_i}$, the Hamiltonian satisfies
\begin{align}
    H_{q_{i}}\left(x^*(t_{s_i}-), \lambda^*(t_{s_i}-), u^*(t_{s_i}-)\right) =H_{q_{i+1}}\left(x^*(t_{s_i}\!+), \lambda^*(t_{s_i}\!+), u^*(t_{s_i}\!+)\right)
\end{align}
\hfill
$\square$
\end{theorem}

The determination of the optimal inputs using Theorem~\ref{thm:HMP} is detailed as follows.

\subsection{Hamiltonian Minimization}\vspace{6pt}

For $i \in \{1,4\}$, which corresponds to the phase-dependent Hamiltonians and cost weights associated with the RTO phases, the Hamiltonians are written as
\begin{equation}
\begin{aligned}
H_{q_{1}}^{i} &= a_{E_i} E + a_{I_i} I + a_{J_i} J + b_{j_i} u_j^2 + \lambda_{V}(-\beta_v V I) + \lambda_{S}(-\beta_s S I + \omega R) + \lambda_{E}(\beta_v V I + \beta_s S I - \kappa E)\\
&\quad + \lambda_{I}(\kappa E - \gamma I - u_jI) + \lambda_{J}(u_jI - \delta J) + \lambda_{R}(\gamma I + \delta J - \omega R).
\end{aligned}
\end{equation}

Since
\begin{equation}\label{q1_opt}
\frac{\partial H_{q{_1}}^{i}}{\partial u}= 0 \Rightarrow  u_j = \frac{(\lambda_I - \lambda_J)I}{2b_{j_i}},
\end{equation}
and given the input constraint $u_j \in [0, u_j^{max}]$, the Hamiltonian minimization \eqref{eq.H} for the RTO phases yields
\begin{equation}
    u_j^* = \min \bigg\{ \max \bigg\{0, \frac{(\lambda_I - \lambda_J)I}{2b_{j_i}} \bigg\} , u_j^{max} \bigg\}
\end{equation}

The Hamiltonian associated with the WFH phase is written as
\begin{equation}
\begin{aligned}
H_{q_2} &= a_{I_2} I + a_{J_2} J + a_{H_{v_2}} H_v + a_{H_{s_2}} H_s + c_2 + b_{j_2} u_j^2 + b_{\sigma_{v_2}} u_{\sigma_v}^2 + b_{\sigma_{s_2}} u_{\sigma_s}^2 + \lambda_{H_v}(u_{\sigma_v} V) + \lambda_{H_s}(u_{\sigma_s} S)\\
&\quad + \lambda_{V}(-\beta_v V I - u_{\sigma_v} V) + \lambda_{S}(-\beta_s S I - u_{\sigma_s} S + \omega R) + \lambda_{E}(\beta_v V I + \beta_s S I - \kappa E)\\
&\quad + \lambda_{I}(\kappa E - \gamma I - u_jI) + \lambda_{J}(u_jI - \delta J) + \lambda_{R}(\gamma I + \delta J - \omega R).
\end{aligned}
\end{equation}

Since
\begin{equation}\label{q2_opt}
\frac{\partial H_{q{_2}}}{\partial u}= 0 \Rightarrow
\begin{cases}
u_j = \dfrac{(\lambda_I - \lambda_J)I}{2b_{j_2}} \\[2mm]
u_{\sigma_v} = \dfrac{(\lambda_{V} - \lambda_{H_v})V}{2b_{\sigma_{v_2}}} \\[2mm]
u_{\sigma_s} = \dfrac{(\lambda_{S} - \lambda_{H_s})S}{2b_{\sigma_{s_2}}}
\end{cases}
\end{equation}
and given the input constraints $u_j \in [0, u_j^{max}]$, $u_{\sigma_v} \in [0, u_{\sigma_v}^{max}]$, and $u_{\sigma_s} \in [0, u_{\sigma_s}^{max}]$, the Hamiltonian minimization \eqref{eq.H} for the WFH phase yields
\begin{equation}
\begin{aligned}
u_j^* &= \min \bigg\{ \max \bigg\{0,\dfrac{(\lambda_I - \lambda_J)I}{2b_{j_2}} \bigg\} ,\; u_j^{max} \bigg\} \\
u_{\sigma_v}^* &= \min \bigg\{ \max\bigg\{ 0,\dfrac{(\lambda_{V} - \lambda_{H_v})V}{2b_{\sigma_{v_2}}} \bigg\},\; u_{\sigma_v}^{max} \bigg\} \\
u_{\sigma_s}^* &= \min \bigg\{\max \bigg\{0, \dfrac{(\lambda_{S} - \lambda_{H_s})S}{2b_{\sigma_{s_2}}} \bigg\},\; u_{\sigma_s}^{max} \bigg\}   
\end{aligned}
\end{equation}

The Hamiltonian associated with the vaccination protocol phase is written as
\begin{equation}
\begin{aligned}
H_{q_3} &= a_{I_3} I + a_{J_3} J + a_{H_{s_3}} H_s + c_3 + b_{j_3} u_j^2 + b_{v_3} u_v^2 + b_{\sigma_{s_3}} u_{\sigma_s}^2 + \lambda_{H_s}(u_{\sigma_s} S - u_v H_s) \\
&\quad + \lambda_{V}(-\beta_v V I + u_v H_s) + \lambda_{S}(-\beta_s S I - u_{\sigma_s} S + \omega R) + \lambda_{E}(\beta_v V I + \beta_s S I - \kappa E) \\
&\quad + \lambda_{I}(\kappa E - \gamma I - u_jI) + \lambda_{J}(u_jI - \delta J) + \lambda_{R}(\gamma I + \delta J - \omega R).
\end{aligned}
\end{equation}
Since
\begin{equation}\label{q3_opt}
\frac{\partial H_{q{_3}}}{\partial u}= 0 \Rightarrow
\begin{cases}
u_j = \dfrac{(\lambda_I - \lambda_J)I}{2b_{j_3}} \\[2mm]
u_v = \dfrac{(\lambda_{H_s} - \lambda_{V})H_s}{2b_{v_3}} \\[2mm]
u_{\sigma_s} = \dfrac{(\lambda_{S} - \lambda_{H_s})S}{2b_{\sigma_{s_3}}}
\end{cases}
\end{equation}
and given the input constraints $u_j \in [0, u_j^{max}]$, $u_v \in [0, u_v^{max}]$, and $u_{\sigma_s} \in [0, u_{\sigma_s}^{max}]$, the Hamiltonian minimization \eqref{eq.H} for the vaccination protocol phase yields
\begin{equation}
\begin{aligned}
u_j^* &= \min \bigg\{\max \bigg\{0, \dfrac{(\lambda_I - \lambda_J)I}{2b_{j_3}} \bigg\},\; u_j^{max} \bigg\} \\[2mm]
u_v^* &= \min \bigg\{\max \bigg\{0, \dfrac{(\lambda_{H_s} - \lambda_{V})H_s}{2b_{v_3}} \bigg\},\; u_v^{max} \bigg\} \\[2mm]
u_{\sigma_s}^* &= \min \bigg\{ \max \bigg\{0,\dfrac{(\lambda_{S} - \lambda_{H_s})S}{2b_{\sigma_{s_3}}} \bigg\},\; u_{\sigma_s}^{max} \bigg\}    
\end{aligned}
\end{equation}
\subsection{Evolution of the Adjoint Process}\vspace{6pt}
For $i \in \{1,4\}$, which corresponds to the phase-dependent adjoint process associated with the RTO phases, the Hamiltonian canonical equation \eqref{eq.la*} in the HMP yields the dynamics of the adjoint process as
\begin{equation} \label{eq:Adjoint14}
\begin{aligned}
\dot{\lambda}_{V}^{i} &=
\beta_v I(\lambda_V - \lambda_E),\\
\dot{\lambda}_{S}^{i} &=
\beta_s I(\lambda_S - \lambda_E),\\
\dot{\lambda}_{E}^{i} &=
-a_{E_i} + \kappa(\lambda_E - \lambda_I),\\
\dot{\lambda}_{I}^{i} &=
-a_{I_i}
+ \beta_v V(\lambda_V - \lambda_E)
+ \beta_s S(\lambda_S - \lambda_E) + (\gamma + u_j^*)\lambda_I
- u_j^*\lambda_J
- \gamma\lambda_R,\\
\dot{\lambda}_{J}^{i} &=
-a_{J_i} + \delta(\lambda_J - \lambda_R),\\
\dot{\lambda}_{R}^{i} &=
\omega(\lambda_R - \lambda_{S}).
\end{aligned}
\end{equation}

For $i=1$, the adjoint dynamics \eqref{eq:Adjoint14} is subject to the boundary conditions
\begin{equation}
 \lambda^{1}_{q_{1}}(t_{s_1}) =  \nabla\xi^\top_{q_1q_2}\lambda_{q_2}(t_{s_1}\!+) + p_1\nabla m_{q_1q_2} = 
\begin{bmatrix}
\lambda_V(t_{s_1}\!+)\\
\lambda_S(t_{s_1}\!+)\\
\lambda_E(t_{s_1}\!+)\\
\lambda_I(t_{s_1}\!+)\\
\lambda_J(t_{s_1}\!+)\\
\lambda_R(t_{s_1}\!+)
\end{bmatrix}
+
p_1
\begin{bmatrix}
0\\
0\\
0\\
1\\
0\\
0
\end{bmatrix}
\end{equation}
Furthermore, the Hamiltonian at the optimal switching time, $t_{s_1}$, satisfies
\begin{equation}
H^1_{q_1}(x,\lambda,u)_{(t_{s_1}-)} = H_{q_2}(x,\lambda,u)_{(t_{s_1}\!+)}
\end{equation}

For $i=4$, the adjoint dynamics \eqref{eq:Adjoint14} is subject to the terminal condition
\begin{equation}
    \lambda^{4}_{q_1}(t_f) = \nabla \Phi = \begin{bmatrix}
0\\
0\\
k_E\\
k_I\\
k_J\\
0
\end{bmatrix}
\end{equation}

The dynamics of the adjoint process for the WFH phase results are
\begin{equation}\label{eq:Adjoint2}
\begin{aligned}
\dot{\lambda}_{H_v} &= -a_{H_{v_2}},\\
\dot{\lambda}_{H_s} &= -a_{H_{s_2}},\\
\dot{\lambda}_{V} &=
\beta_v I(\lambda_V-\lambda_E)
+u_{\sigma_v}^{*}(\lambda_V-\lambda_{H_v}),\\
\dot{\lambda}_{S} &=
\beta_s I(\lambda_S-\lambda_E)
+u_{\sigma_s}^{*}(\lambda_S-\lambda_{H_s}),\\
\dot{\lambda}_{E} &=
\kappa(\lambda_E-\lambda_I),\\
\dot{\lambda}_{I} &=
-a_{I_2}
+\beta_v V(\lambda_V-\lambda_E)
+\beta_s S(\lambda_S-\lambda_E) +( \gamma+u_j^{*} )\lambda_I
-u_j^{*}\lambda_J
-\gamma\lambda_R,\\
\dot{\lambda}_{J} &=
-a_{J_2}+\delta(\lambda_J-\lambda_R),\\
\dot{\lambda}_{R} &=
\omega(\lambda_R-\lambda_S).
\end{aligned}
\end{equation}

The adjoint dynamics \eqref{eq:Adjoint2} is subject to the boundary conditions
\begin{equation}
\lambda_{q_2}(t_{s_2}) = \nabla\xi^\top_{q_2q_3}\lambda_{q_3}(t_{s_2}\!+) =
\begin{bmatrix}
\lambda_{V}(t_{s_2}\!+)\\
\lambda_{H_s}(t_{s_2}\!+)\\
\lambda_V(t_{s_2}\!+)\\
\lambda_S(t_{s_2}\!+)\\
\lambda_E(t_{s_2}\!+)\\
\lambda_I(t_{s_2}\!+)\\
\lambda_J(t_{s_2}\!+)\\
\lambda_R(t_{s_2}\!+)
\end{bmatrix}
\end{equation}

Furthermore, the Hamiltonian at the optimal switching time, $t_{s_2}$, satisfies
\begin{equation}
H_{q_2}(x,\lambda,u)_{(t_{s_2}-)} = H_{q_3}(x,\lambda,u)_{(t_{s_2}\!+)}
\end{equation}

The dynamics of the adjoint process for the vaccination protocol phase results are
\begin{equation}\label{eq:Adjoint3}
\begin{aligned}
\dot{\lambda}_{H_s} &=
-a_{H_{s_3}} + u^*_v(\lambda_{H_s} - \lambda_{V}),\\
\dot{\lambda}_{V} &=
\beta_v I(\lambda_{V} - \lambda_E),\\
\dot{\lambda}_{S} &=
\beta_s I(\lambda_{S} - \lambda_E)
+ u^*_{\sigma_s}(\lambda_{S} - \lambda_{H_s}),\\
\dot{\lambda}_{E} &=
\kappa(\lambda_E - \lambda_I),\\
\dot{\lambda}_{I} &=
-a_{I_3}
+ \beta_v V(\lambda_{V} - \lambda_E)
+ \beta_s S(\lambda_{S} - \lambda_E) + (\gamma + u^*_j)\lambda_I
- u^*_j\lambda_J
- \gamma\lambda_R,\\
\dot{\lambda}_{J} &=
-a_{J_3} + \delta(\lambda_J - \lambda_R),\\
\dot{\lambda}_{R} &=
\omega(\lambda_R - \lambda_{S}).
\end{aligned}
\end{equation}

The adjoint dynamics \eqref{eq:Adjoint3} is subject to the boundary conditions
\begin{equation}
 \lambda_{q_{3}}(t_{s_3}) =  \nabla\xi^\top_{q_3q_1}\lambda_{q_1}(t_{s_3}\!+) + p_3\nabla m_{q_3q_1} = 
\begin{bmatrix}
\lambda_{S}(t_{s_3}\!+)\\
\lambda_V(t_{s_3}\!+)\\
\lambda_S(t_{s_3}\!+)\\
\lambda_E(t_{s_3}\!+)\\
\lambda_I(t_{s_3}\!+)\\
\lambda_J(t_{s_3}\!+)\\
\lambda_R(t_{s_3}\!+)
\end{bmatrix}
+
p_3
\begin{bmatrix}
0\\
0\\
0\\
0\\
1\\
0\\
0
\end{bmatrix}
\end{equation} 

Furthermore, the Hamiltonian at the optimal switching time, $t_{s_3}$, satisfies
\begin{equation}
H_{q_3}(x,\lambda,u)_{(t_{s_3}-)} = H^4_{q_1}(x,\lambda,u)_{(t_{s_3}\!+)}
\end{equation}

\section{Numerical Simulation}\vspace{6pt}
\label{sec:num_sim}
We consider the phase-dependent cost weights provided in Table~\ref{tab:params_table}, together with the initial condition
$x_{q_1}(t_0) = [0.10,\,0.74,\,0.05,\,0.01,\,0.05,\,0.05]^\top$,
which corresponds to a population that is predominantly susceptible and unvaccinated, with a small fraction of exposed, infectious, quarantined, and recovered individuals initiating the outbreak. For all phases, we consider $\beta_{v}=0.18$, $\beta_{s}=0.30$, $\kappa=0.20$, $\gamma=0.13$, $\delta=0.18$, and $\omega=0.02$.

The hybrid optimal control problem is solved over the finite time horizon $[t_0,t_f] = [0,40]$ days. The optimal state trajectories, control inputs, adjoint variables, and Hamiltonians are computed using a direct transcription approach via the GPOPS-II software package \citep{GPOPS}.

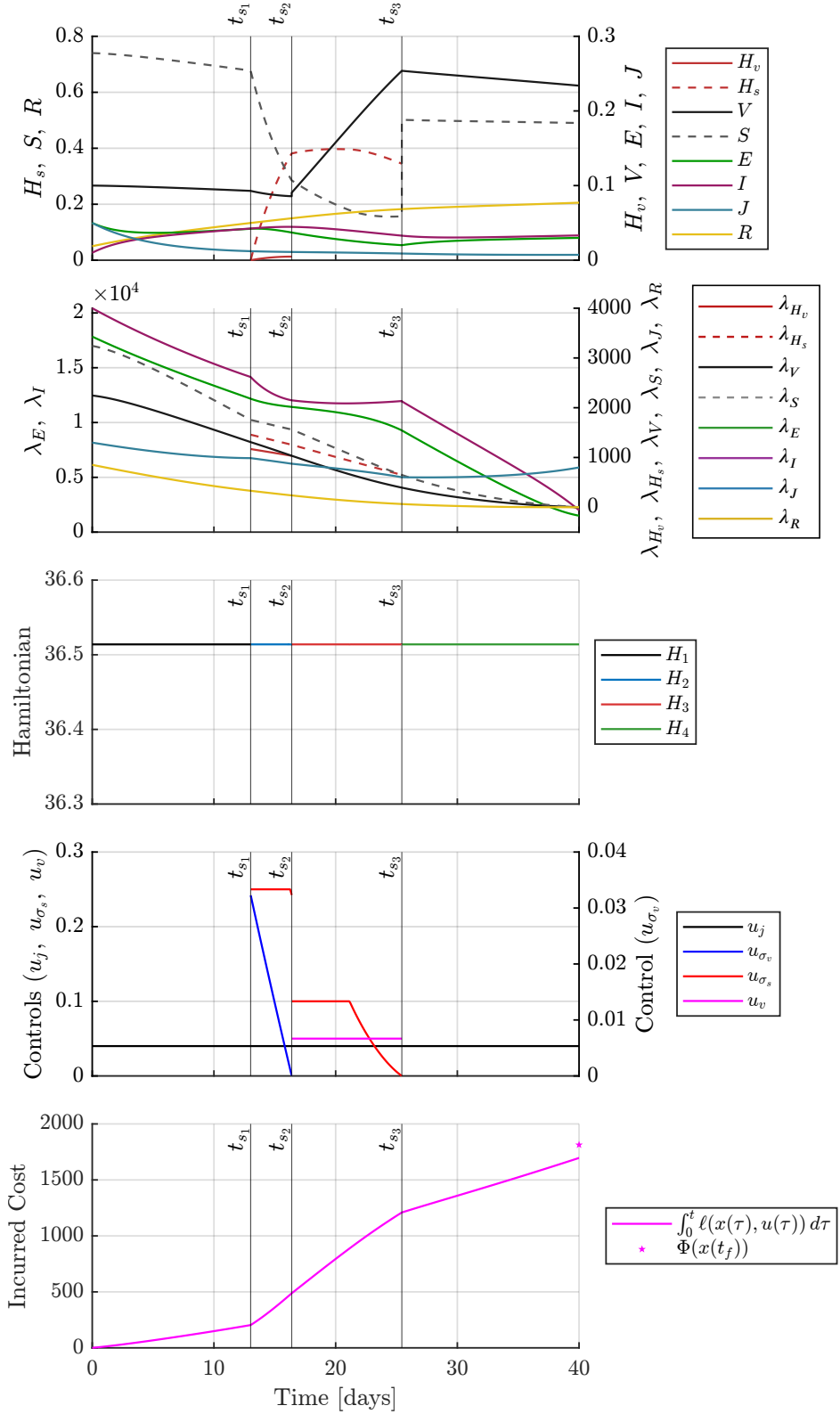
\begin{figure}[H]
\centering
\begin{overpic}[width=0.75\textwidth]{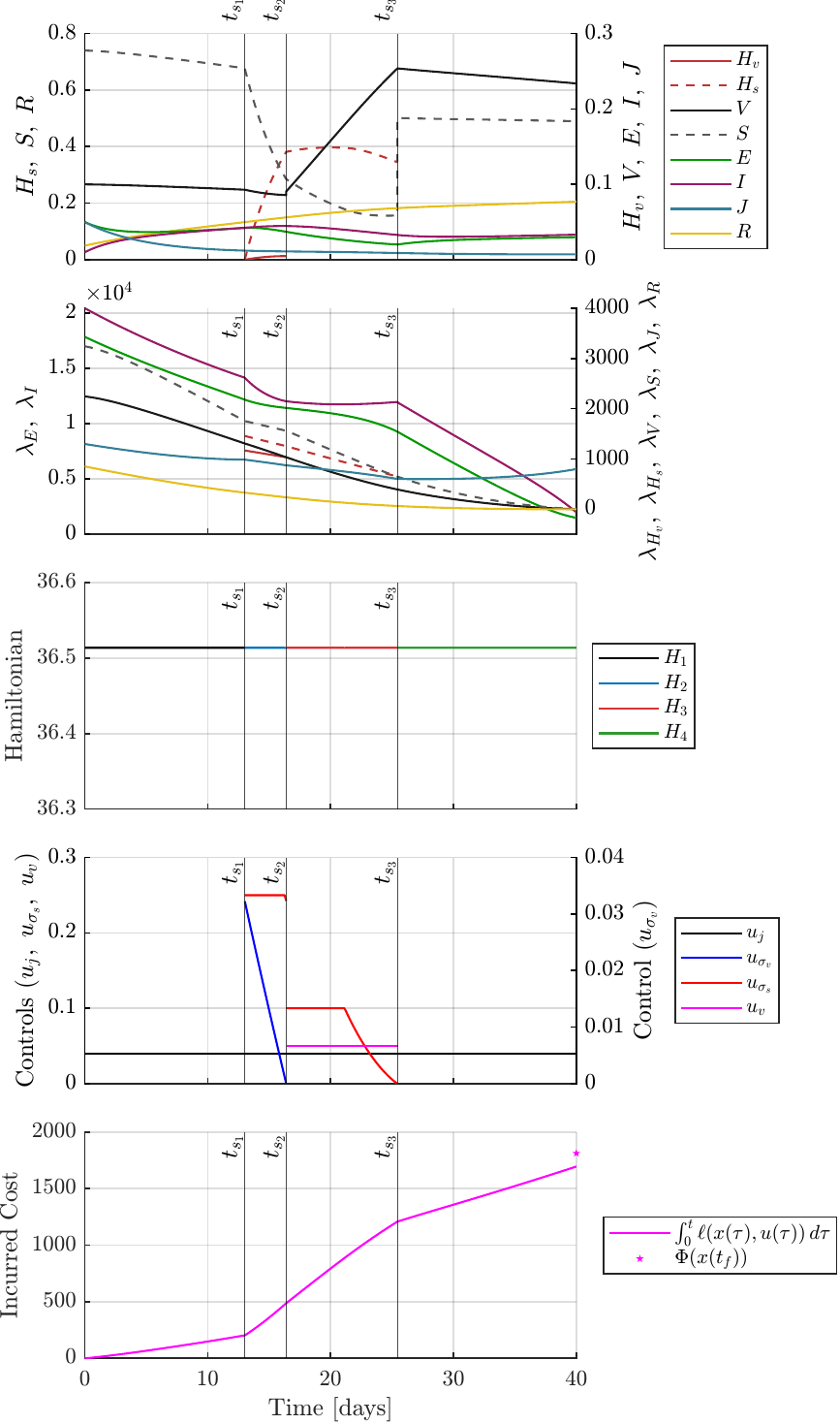}

\put(48,62){ 

\begin{tikzpicture}
\node[draw, fill=white, inner sep=2pt, line width=0.6pt] {

\begin{tikzpicture}[x=1cm,y=0.45cm]

\draw[HvColor, line width=0.9pt] (0,7) -- (1.1,7);
\node[right] at (1.1,7) {$\lambda_{H_v}$};

\draw[HvColor, dashed, line width=0.9pt] (0,6) -- (1.1,6);
\node[right] at (1.1,6) {$\lambda_{H_s}$};

\draw[VColor, line width=0.9pt] (0,5) -- (1.1,5);
\node[right] at (1.1,5) {$\lambda_{V}$};

\draw[SColor, dashed, line width=0.9pt] (0,4) -- (1.1,4);
\node[right] at (1.1,4) {$\lambda_{S}$};

\draw[EColor, line width=0.9pt] (0,3) -- (1.1,3);
\node[right] at (1.1,3) {$\lambda_{E}$};

\draw[IColor, line width=0.9pt] (0,2) -- (1.1,2);
\node[right] at (1.1,2) {$\lambda_{I}$};

\draw[JColor, line width=0.9pt] (0,1) -- (1.1,1);
\node[right] at (1.1,1) {$\lambda_{J}$};

\draw[RColor, line width=0.9pt] (0,0) -- (1.1,0);
\node[right] at (1.1,0) {$\lambda_{R}$};

\end{tikzpicture}

};
\end{tikzpicture}

}

\end{overpic}

\caption{Optimal states, adjoints, Hamiltonian, controls, and incurred cost.}
\label{fig:results}
\end{figure}

The resulting optimal control strategy, shown in Fig.~\ref{fig:results}, yields optimal switching times
$[t_{s_1},\,t_{s_2},\,t_{s_3}] = [13.02,\,16.39,$\\$\,25.44]$ \;\text{days},
where $t_{s_1}$ denotes the transition from the first RTO phase to the WFH phase, $t_{s_2}$ corresponds to the activation of the vaccination protocol phase, and $t_{s_3}$ represents the return to the final RTO phase. The associated optimal switching states are given by $x_{q_1}(t_{s_1}-)=[0.093,\,0.677,\,0.042,\,0.043,\,0.012,\,0.133]^\top$,
$x_{q_2}(t_{s_2}-)=[0.005,\,0.382,\,0.086,\,0.285,\,0.037,\,0.045,\,0.011,\,0.150]^\top$, 
and $x_{q_3}(t_{s_3}-)=[0.345,\,0.254,\,0.157,\,0.020,\,0.033,\,0.009,\,$\\$0.183]^\top$, which capture the system states immediately prior to each phase transition. These switching states reflect the evolving balance between susceptible, infected, and quarantined populations as intervention strategies are introduced and withdrawn. The epidemiological system terminates at $t_f$ with $x_{q_1}(t_{f})=[0.234,\,0.490,\,0.030,\,0.033,\,0.007,\\\,0.206]^\top$.  

\begin{figure}[H]
    \centering
    \begin{tikzpicture}
        \node[inner sep=0] (img) at (0,0) {
            \includegraphics[width=\linewidth,height=0.88\textheight,keepaspectratio]{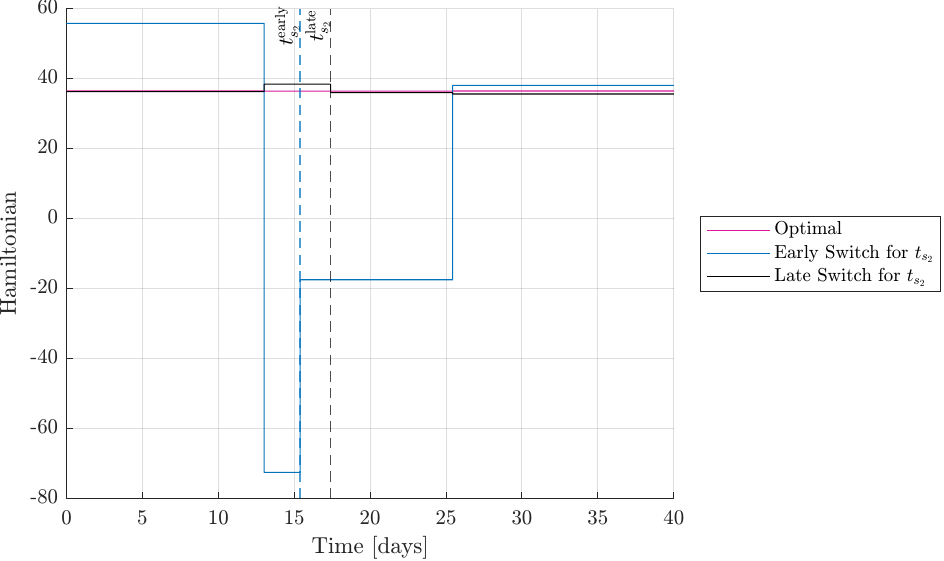}
        };
        \begin{scope}[
            x={($ (img.south east) - (img.south west) $)},
            y={($ (img.north west) - (img.south west) $)},
            shift={(img.south west)}
        ]
            \coordinate (x_start) at (0.09, 0); 
            \coordinate (x_end)   at (0.72, 0); 
            
            \coordinate (y_start) at (0, 0.11);
            \coordinate (y_end)   at (0, 0.837);
            \coordinate (t_val) at ($ (x_start) ! 0.387 ! (x_end) $);

            \draw[magenta, thick, dash pattern=on 5pt off 3pt] 
                (t_val |- y_start) -- (t_val |- y_end)
                node[pos=1.05, text=magenta, rotate=90,  font=\normalsize] {$t_{s_2}^*$};
                
        \end{scope}
    \end{tikzpicture}
    \caption{Hamiltonian comparison for non-optimal switching times.}
    \label{fig:hamiltonian_compare}
\end{figure}
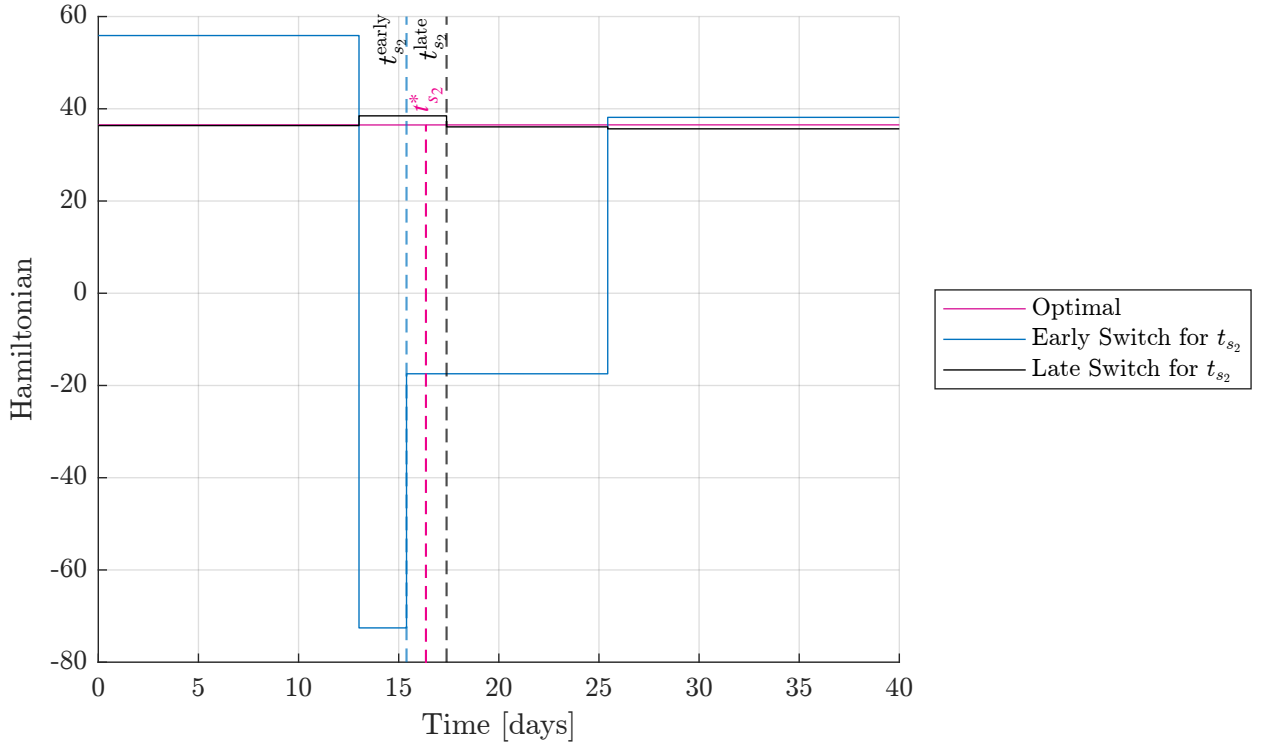

\begin{figure}[H]
\centering
\includegraphics[width=\linewidth,height=0.88\textheight,keepaspectratio]{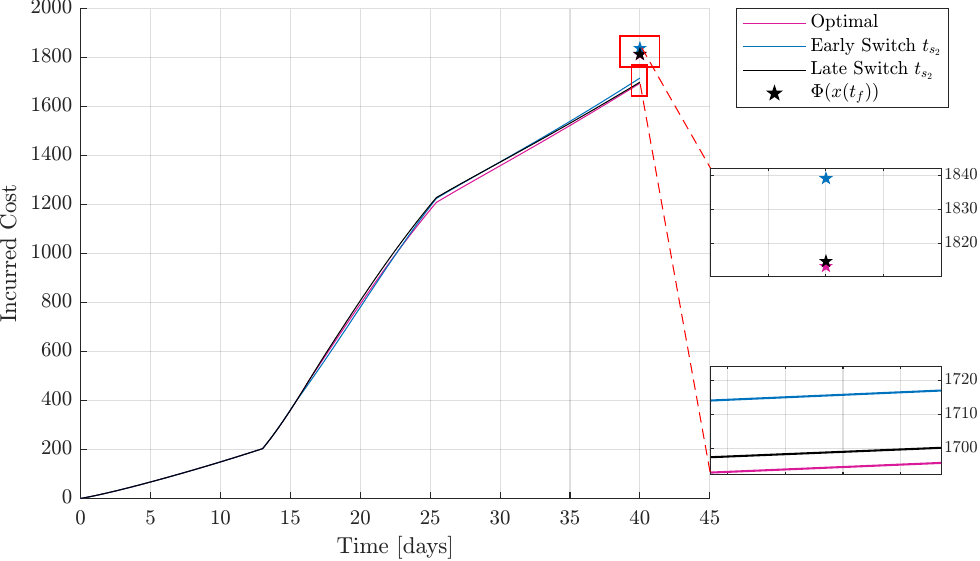}\\
\caption{Incurred cost comparison for non-optimal switching times.}
\label{fig:zoomed_cost}
\end{figure}

To further assess the hybrid optimal control framework, the controlled switching time $t_{s_2}$ is perturbed from its optimal value by $\pm 1$ day to simulate non-optimal switching configurations. As confirmed by the minimum total cost achieved in Fig.~\ref{fig:zoomed_cost} (where a star indicates the inclusion of terminal costs), the suboptimality of the perturbed schedules is mathematically indicated by the loss of Hamiltonian continuity shown in Fig.~\ref{fig:hamiltonian_compare}—a violation of HMP requirements that the HMP-MAS algorithm in \citep{PakniyatCaines2021} leverages to update non-optimal switching times until it eventually yields the optimal inputs and minimum total cost. 

\FloatBarrier

\begin{table}[width=.9\linewidth,cols=5,pos=h]
\caption{Epidemiological Phase-Dependent Weights}
\label{tab:params_table}

\begin{tabular*}{\tblwidth}{@{}LCCCC@{}}
\toprule
\textbf{Phase} & $\ell_{q_1}^{1}$ & $\ell_{q_2}$ & $\ell_{q_3}$ & $\ell_{q_1}^{4}$ \\
\midrule

$a_I$ & 400 & 1200 & 1100 & 950 \\
$a_J$ & 120 & 150 & 140 & 100 \\
$a_E$ & -- & -- & -- & 100 \\
$a_{H_v}$ & -- & 40 & -- & -- \\
$a_{H_s}$ & -- & 60 & 80 & -- \\

\midrule

$b_j$ & 100 & 100 & 100 & 60 \\
$b_{\sigma_v}$ & -- & 200 & -- & -- \\
$b_{\sigma_s}$ & -- & 180 & 120 & -- \\
$b_v$ & -- & -- & 160 & -- \\
$c$ & -- & 5.00 & 2.50 & -- \\

\midrule

$k_E$ & -- & -- & -- & 1500 \\
$k_I$ & -- & -- & -- & 2000 \\
$k_J$ & -- & -- & -- & 800 \\

\midrule

$I_{\text{high}}$ & 0.043 & -- & -- & -- \\
$I_{\text{low}}$ & -- & -- & -- & 0.033 \\

\bottomrule
\end{tabular*}
\end{table}
\FloatBarrier
\section{Conclusion}\vspace{6pt}
\label{sec:conclusion}
This paper presents a hybrid optimal control framework for multi-phase epidemiological systems with regime-switching dynamics and time-dependent interventions. The formulation captures structural changes in system dynamics through phase-dependent models and running costs that reflect evolving policy objectives. Using the Hybrid Minimum Principle (HMP), the framework jointly determines optimal controls and switching times, balancing intervention effort and epidemic burden. Results show that combining WFH policies with vaccination protocol improves performance over single-phase strategies. Future work will extend this framework to heterogeneous, network-based settings as well as large-scale populations with non-uniform structured interactions and their associated graphon limits.

\printcredits

\section*{Declaration of competing interest}\vspace{6pt}

The authors declare that they have no known competing financial interests or personal relationships that could have appeared to influence the work reported in this paper.

\section*{Data availability}\vspace{6pt}

No data was used for the research described in the article.
\bibliographystyle{elsarticle-num}
\bibliography{cas-refs}\vspace{6pt}

\end{document}